\documentclass{llncs}
\usepackage{float}

\usepackage[usenames,dvipsnames]{color}

\usepackage{wrapfig}

\usepackage{url}
\usepackage{eurosym}
\usepackage{fancyvrb}

\usepackage{amssymb}

\usepackage{url}

\usepackage{graphicx}              
\usepackage[latin2]{inputenc}
\usepackage{mathtools}
\usepackage{ifthen}
\usepackage{array}
\usepackage{tikz}
\usetikzlibrary{decorations.pathreplacing,calc}
\usepackage{verbatim}
\usepackage{arydshln}
\usepackage{float}
\usepackage{array}
\usepackage{hyperref}

\usepackage{times}

\newcommand{\draw}{\leftarrow}

\newcommand{\pk}{\mathit{pk}}
\newcommand{\sk}{\mathit{sk}}

\newcommand{\Dec}{\textrm{Dec}}

\newcommand{\Enc}{\textrm{Enc}}

\DeclareFontShape{OT1}{cmtt}{bx}{n}{<5><6><7><8><9><10><10.95><12><14.4><17.28><20.74><24.88>cmttb10}{}

\newcommand{\tT}{\hat{T}}

\newcommand{\true}{\mathrm{true}}
\newcommand{\false}{\mathrm{false}}
\newcommand{\inp}{\textbf{in:}}

\newcommand{\argu}{\mathit{arg}}
\newcommand{\BB}{\mathsf{BlockChain}}
\newcommand{\maxBB}{\mathsf{max_{BB}}}
\newcommand{\BODY}[1]{[#1]}

\newcommand{\uX}{\mathsf{X}}
\newcommand{\uY}{\mathsf{Y}}

\newcommand{\Open}{\mathit{Open}}
\newcommand{\Fuse}{\mathit{Fuse}}

\newcommand{\sP}{\mathsf{P}}

\newcommand{\sA}{\mathsf{A}}
\newcommand{\sB}{\mathsf{B}}

\newcommand{\graphC}[1]{\begin{center}\graph{#1}\end{center}}

\newcommand{\out}[1]{}

\newcommand{\up}{\vspace{-0.1truecm}}

\newcommand{\UPPB}[1]{$\texttt{\textbf{#1}}$}
\newcommand{\UPP}[1]{${\tt #1}$}

\newcommand{\Drect}[4]{
  \node[rectangle] (#3) at (#1) 
  {\begin{tabular}{|c|} \hline \\ \hspace{#2 cm} #4 \hspace{#2 cm} \\ \\ \hline \end{tabular}};
}

\newcommand{\NSCS}{{\mathsf{NewSCS}}}
\newcommand{\NSCSCom}{{\mathsf{NewSCS.Commit}}}
\newcommand{\NSCSOp}{{\mathsf{NewSCS.Open}}}
\newcommand{\CSCom}{{\mathsf{CS.Commit}}}
\newcommand{\CSOp}{{\mathsf{CS.Open}}}
\newcommand{\ComS}{{\mathsf{CS}}}

\tikzset{>=latex}

\newcommand{\sig}{\mathsf{sig}}
\newcommand{\ver}{\mathsf{ver}}

\newcommand{\Commit}{\mathit{Commit}}
\newcommand{\Reveal}{\mathit{Open}}
\newcommand{\FuseAR}{\mathit{Fuse}}
\newcommand{\CommitSCS}{\mathit{Commit}}
\newcommand{\RevealSCS}{\mathit{Open}}
\newcommand{\FuseSCS}{\mathit{Fuse}}

\newcommand{\graph}[1]
{
  {
  
    \footnotesize
    \centering
    \begin{tikzpicture}[
      nodes={
	inner sep=-0.5\pgflinewidth, outer sep=-0.5\pgflinewidth,
	inner xsep=-3.5\tabcolsep, 
	outer xsep=0.5\pgflinewidth, 
	align=left
      }]
      #1
    \end{tikzpicture}
    
  }
}

\newcommand{\desc}[2][0.95]
{
  \centering
  \fbox{ 
    \footnotesize
    \begin{minipage}[c]{#1\linewidth}
      #2
    \end{minipage}
  }
}

\newcommand{\transactionAux}[7]{
  \begin{tabular}{|m{#2 cm}|}
    \hline
    \multicolumn{1}{|c|}{\textbf{#3}} \\
    \hline
    #4 \\
    \hline
    \textbf{out-script(#5):} #6 \\
    \textbf{val:} #7
    \ifthenelse{\not \equal{#1}{0}}
      {\\ \hline
	\textbf{tlock:} #1 \\
	\hline}
      {\\ \hline}
  \end{tabular}
}

\newcommand{\DtransactionAux}[7]{
  \begin{tabular}{:m{#2 cm}:}
    \hdashline
    \multicolumn{1}{:c:}{\textbf{#3}} \\
    \hdashline
    #4 \\
    \hdashline
    \textbf{out-script(#5):} #6 \\
    \textbf{val:} #7
    \ifthenelse{\not \equal{#1}{0}}
      {\\ \hdashline
	\textbf{tlock:} #1 \\
	\hdashline}
      {\\ \hdashline}
  \end{tabular}
}

\newcommand{\DtransTwoInTwoOut}[9]{
  \node[rectangle] (#3) at (#1)
  {\DtransactionAuxTwoInTwoOut{#2}{#4}{#5}{#6}{#7}{#8}{#9}};
}

\newcommand{\DtransactionAuxTwoInTwoOut}[7]{
  \begin{tabular}{|m{#1 cm}|}
    \hline
    \multicolumn{1}{|c|}{\textbf{#2}} \\
    \hline
    \begin{tabular}{m{0.48\linewidth} | m{0.47\linewidth}}
        #3
    \end{tabular}
    \\
    \hline
    \begin{tabular}{m{0.48\linewidth} | m{0.47\linewidth}}
      \textbf{out-script$_1$(#4):} #5 &
      \textbf{out-script$_2$(#4):} #6 \\
      \textbf{val$_1$:} #7 & \textbf{val$_2$:} #7
    \end{tabular}
    \\
    \hline
  \end{tabular}
}

\newcommand{\transaction}[7][0]{
  \transactionAux{#1}{#2}{#3}{#4}{#5}{#6}{#7}
}

\newcommand{\Dtransaction}[7][0]{
  \DtransactionAux{#1}{#2}{#3}{\textbf{in-script:} #4}{#5}{#6}{#7}
}

\newcommand{\trans}[9][0]{
  \node[rectangle] (#4) at (#2)
  {\transaction[#1]{#3}{#5}{\textbf{in-script:} #6}{#7}{#8}{#9}};
}

\newcommand{\Dtrans}[9][0]{
  \node[rectangle] (#4) at (#2)
  {\Dtransaction[#1]{#3}{#5}{#6}{#7}{#8}{#9}};
}

\newcommand{\transTwoInWithScripts}[9][0]{
  \node[rectangle,] (#4) at (#2)
  {
    \transaction[#1]{#3}{#5}{
      \hspace{-0.12cm}
      \begin{tabular}{p{0.48\linewidth}|p{0.47\linewidth}}
	#6
      \end{tabular}
    }{#7}{#8}{#9}};
}

\newcommand{\inScripts}[2]{
  \textbf{in-script:} #1  &
  \textbf{in-script:} #2
}

\makeatletter
\newcommand{\thickhline}{%
    \noalign {\ifnum 0=`}\fi \hrule height 1pt
    \futurelet \reserved@a \@xhline
}
\newcolumntype{"}{@{\hskip\tabcolsep\vrule width 1pt\hskip\tabcolsep}}
\makeatother

\newcommand{\BTC}{%
  \leavevmode
  \vtop{\offinterlineskip 
    \setbox0=\hbox{B}%
    \setbox2=\hbox to\wd0{\hfil\hskip-.03em
    \vrule height .3ex width .15ex\hskip .08em
    \vrule height .3ex width .15ex\hfil}
    \vbox{\copy2\box0}\box2}}

\newcommand{\AR}{{\mathsf{CS}}}

\newcommand{\BCA}{{\UPP{Block}\-\UPP{Chain}\-\UPP{Agent}}}
\newcommand{\BC}{{\UPP{bc}}}
\newcommand{\Helper}{{\UPP{Helper}}}
\newcommand{\UNSENT}{{\UPP{UNSENT}}}
\newcommand{\SENT}{{\UPP{SENT}}}
\newcommand{\CONFIRMED}{{\UPP{CONFIRMED}}}
\newcommand{\SPENT}{{\UPP{SPENT}}}
\newcommand{\CANCELLED}{{\UPP{CANCELLED}}}
\newcommand{\LATENCY}{{\UPP{MAX\_LATENCY}}}
\newcommand{\CS}{{\UPP{CS}}}
\newcommand{\NewSCS}{{\mathsf{NewSCS}}}

\newcommand{\kA}{\mathcal{A}}
\newcommand{\kB}{\mathcal{B}}

\usepackage{times}

\usepackage{framed}
\newcommand{\UPPAAL}{{\sc Uppaal}}

\usepackage[T1]{fontenc}

\begin{document}

\title{Modeling Bitcoin Contracts by Timed Automata\thanks{This work was supported by the WELCOME/2010-4/2 grant founded within the framework of the EU Innovative Economy (National Cohesion Strategy) Operational Programme.}}

\author{Marcin Andrychowicz, Stefan Dziembowski\thanks{On leave from the {\em Sapienza} University of Rome.}, Daniel Malinowski and {\L}ukasz Mazurek}

\institute{Cryptology and Data Security Group\\ \url{www.crypto.edu.pl}\\University of Warsaw}

\maketitle


\begin{abstract}
Bitcoin is a peer-to-peer cryptographic currency system. Since its introduction in 2008, Bitcoin has gained noticeable popularity, mostly due to its following properties: (1)   the transaction fees are very low, and (2) it is not controlled by any central authority, which in particular means that nobody can ``print'' the money to generate inflation. Moreover, the transaction syntax allows to create the so-called {\em contracts}, where a number of mutually-distrusting parties engage in a protocol to jointly perform some financial task, and the fairness of this process is guaranteed by the properties of Bitcoin.  Although the Bitcoin contracts have several potential applications in the digital economy, so far they have not been widely used in real life. This is partly due to the fact that they are cumbersome to create and analyze, and hence risky to use. 

In this paper we propose to remedy this problem by using the methods originally developed for the computer-aided analysis for hardware and software systems, in particular those based on the timed automata.  More concretely, we propose a  framework for modeling the Bitcoin contracts using the timed automata in the \UPPAAL~model checker. Our method is general and can be used to model several contracts.  As a proof-of-concept we  use this framework to model some of the Bitcoin contracts from our recent previous work.  We then automatically verify their security in \UPPAAL, finding (and correcting) some subtle errors that were difficult to spot by the manual analysis. We hope that our work can draw the attention of the 
researchers working on formal modeling to the problem of the Bitcoin contract verification, and spark off more research on this topic.
\end{abstract}


\section{Introduction}

Bitcoin is a digital currency system introduced in 2008 by an anonymous developer using a pseudonym ``Satoshi Nakamoto'' \cite{nakamoto}.  Despite of its mysterious origins, Bitcoin became the first cryptographic currency that got widely adopted --- as of January 2014 the Bitcoin capitalization is over \EUR{7 bln}.  The enormous success of Bitcoin was also widely covered by the media (see e.g.~\cite{economist1,NYT,Forbes,WP,CNN}) and even attracted the attention of several governing bodies and legislatures, including the US Senate \cite{WP}. Bitcoin owes its popularity mostly to the fact that it has no central authority, the transaction fees are very low, and the amount of coins in the circulation is restricted, which in particular means that nobody can ``print'' money to generate inflation. The financial transactions between the participants are published on a public ledger maintained jointly by the users of the system.  

One of the very interesting, but slightly less known, features of the Bitcoin is the fact that it allows for more complicated ``transactions'' than the simple money transfers between the participants: very informally, in Bitcoin it is possible to ``deposit'' some amount of money in such a way that it can be claimed only under certain conditions.  These conditions  are written in the form of the {\em Bitcoin scripts} and in particular may involve some timing constrains.  This property allows to create the so-called {\em contracts} \cite{Contr}, where a number of mutually-distrusting parties engage in a Bitcoin-based protocol to jointly perform some task.
The  security of the protocol is guaranteed purely by the properties of the Bitcoin, and no additional trust assumptions are needed.  This Bitcoin feature can have  several applications in the digital economy, like creating the assurance contracts, the escrow and dispute mediation, the rapid micropayments \cite{Contr},  the multiparty lotteries \cite{ADMM13}. It can also be used to add some extra properties to Bitcoin, like the certification of the users \cite{cryptoeprint:2014:076}, or creating the secure ``mixers'' whose goal is to enhance the anonymity of the transactions \cite{boyen}.  Their potential has even been noticed by the  media (see, e.g., a recent enthusiastic article on the {\em CNN Money} \cite{CNN}). 

In our opinion, one of the obstacles that may prevent this feature from being widely used by the Bitcoin community is the fact that the contracts are tricky to write and understand. This may actually be the reason why, despite of so many potential applications, they have not been widely used in real life. As experienced by ourselves \cite{ADMM13b,ADMM13c,ADMM13}, developing such contracts is hard for the following reasons.  Firstly, it's easy to make subtle mistakes in the scripts.  Secondly, the protocols that involve several parties and the timing constraints are naturally hard to analyze by hand.  Since  mistakes in the contracts can be exploited by the malicious parties for their own financial gain, it is natural that users are currently reluctant to use this feature of Bitcoin.  

In this paper we propose an approach that can help  designing secure Bitcoin contracts. Our idea is to use the methods originally developed for the computer-aided analysis for hardware and software systems, in particular the timed automata \cite{icalp1990-AD,Alur94atheory}. They seem to be the right tool for this purpose due to the fact that the protocols used in the Bitcoin contracts typically have a finite number of states and depend on the notion of time. This time-dependence is actually two-fold, as (1) it takes some time for the Bitcoin transactions to be confirmed (1 hour, say), and (2) the Bitcoin transactions can come with a ``time lock'' which specifies the time when a transaction becomes valid.  

\vspace{-.5truecm}

\subsubsection{Our contribution}
We propose a  framework for modeling the Bitcoin contracts using timed automata in the \UPPAAL~model checker \cite{UPPAAL,UPPAAL2} (this is described in Sec.~\ref{sec:mod}).
Our method is general and can be used to model a wide class of contracts.
As a proof-of-concept, in Sec.~\ref{sec:commit} we  use this framework  to model two Bitcoin contracts from our previous work \cite{ADMM13,ADMM13b}.
This is done manually, but our method is quite generic and can potentially be automatized. 
In particular, most of the code in our implementation does not depend on the protocol being verified, but describes the properties of Bitcoin system.
To model a new contract it is enough to specify the transactions used in the contract, the knowledge of the parties at the beginning of the protocol
and the protocol followed by the parties. 
We then automatically verify the security of our contracts in \UPPAAL{} (in Sec.~\ref{sec:res}).
The \UPPAAL{} code for the contracts modeled and verified by us is available at the web page \url{http://crypto.edu.pl/uppaal-btc.zip}.


\vspace{-.3truecm}

\subsubsection{Future work}\label{sec:future}

We hope that our work can draw the attention of the researchers
working on formal modeling to the problem of the Bitcoin contracts verification, and spark off more research on this topic.  What seems especially interesting is to try to fully automatize this process. One attractive option is to think of the following workflow: (1) a designer of a Bitcoin contract describes it in \UPPAAL{} (or, possibly, in some extension of it), (2) he verifies the security of this idealized description using \UPPAAL{}, and (3) if the idealized description verifies correctly, then he uses the system to ``compile'' it into a real Bitcoin implementation that can be deployed in the wild.  Another option would be to construct a special tool for designing the Bitcoin contracts, that would produce two outputs: (a) a code in the \UPPAAL{} language (for verification) and (b) a real-life Bitcoin implementation. 

Of course, in both cases one would need to formally show the soundness of this process (in particular: that the ``compiled'' code maintains the properties of the idealized description). Hence, this project would probably require both non-trivial theoretical and engineering work. 


\vspace{-.3truecm}

\subsubsection{Preliminaries}

Timed automata were introduced by Alur and Dill \cite{icalp1990-AD,Alur94atheory}. There exist other model checkers based on this theory, like Kronos \cite{Kronos} and Times \cite{Times}. It would be interesting to try to implement our ideas also in them.  Other formal models that involve the notion of the real time include the timed Petri nets \cite{Petri}, the timed CSP \cite{CSP}, the timed process algebras \cite{PA,NicollinS94}, and the timed propositional temporal logic \cite{jacm41(1)-AH}. One can try to model the Bitcoin contracts also using these formalisms.  For the lack of space, a short introduction to UPPAAL was moved to Appendix~\ref{sec:uppaal-desc}. The reader may also consult \cite{UPPAAL,UPPAAL2} for more information on this system.

We assume reader's familiarity with the public-key cryptography, in particular with the signature schemes (an introduction to this concept can be found e.g.~in \cite{KL,DelfsKnebl}).  
We will frequently denote the key pairs using the capital letters (e.g.~$A$), 
and refer to the private key and the public key of $A$ by: $A.\sk$ and $A.\pk$,
respectively.  
We will also use the following convention: if $A=(A.\sk,A.\pk)$ then  $\sig_{A}(m)$
denotes a signature  on a message $m$ computed with $A.\sk$ and  
$\ver_{A}(m,\sigma)$ denotes the result ($\true$ or $\false$) of the verification 
of a signature $\sigma$ on message $m$ with respect to the public key $A.\pk$.  We will use the ``$\BTC$'' symbol to denote the Bitcoin currency unit.


\renewcommand{\up}{}
\up\up\up

\subsection{A short description of Bitcoin}\label{sec:bc-desc}

\up

Since we want the exposition to be self-contained, we start with a short description of Bitcoin, focusing only on the most relevant parts. For the lack of space we do not describe how the coins are created, how the transaction fees are charged, and how the Bitcoin ``ledger'' is maintained. A more detailed description of Bitcoin is available on the {\em Bitcoin wiki} site \cite{wiki}.  The reader may also consult the original Nakamoto's paper \cite{nakamoto}.

\vspace{-.3truecm}

\subsubsection{Introduction}
In general one of the main challenges when designing a digital currency is the potential double spending: if coins are just strings of bits then the owner of a coin can spend it multiple times.  Clearly this risk could be avoided if the users have access to a trusted ledger with the list of all the transactions.  In this case a transaction would be considered valid only if it is posted on the board.  For example suppose the transactions are of a form: ``user $\uX$ transfers to user $\uY$ the money that he got in some previous transaction $T_{p}$'', signed by the user $\uX$.
In this case each user can verify if money from transaction $T_{p}$ has not been already spent by $\uX$.  The main difficulty in designing the fully-distributed peer-to-peer currency systems is to devise a system where the users jointly maintain the ledger in such a way that it cannot be manipulated by an adversary and it is publicly-accessible. 

In Bitcoin this problem is solved by a cryptographic tool called {\em proofs-of-work} \cite{C:DwoNao92}.   We will not go into the details of how this is done, since it is not relevant to this work.
%
 Let us only say that the system works securely as long as no adversary controls more  computing power than the combined computing power of all the other participants of the protocol\footnote{It is currently estimated \cite{Forbes} that the combined computing power of the Bitcoin participants is around 64 exaFLOPS, which exceeds by factor over 200 the total computing power of world's top 500 supercomputers, hence the cost of purchasing the equipment that would be needed to break this system is huge.}. The Bitcoin participants that contribute their computing power to the system are called the {\em miners}. Bitcoin contains a system of incentives to become a miner. For the lack of space we do not describe it here.
 
%
%
%

Technically, the ledger is implemented as a chain of blocks, hence it is also called a ``block chain''.  When a transaction is posted on the block chain, it can take some time before it appears on it, and even some more time before the user can be sure that this transaction will not be cancelled. However, it is safe to assume that there exists an upper bound on this waiting time (1-2 hours, say). We will denote this time by \LATENCY.
  
As already highlighted in the introduction, the format of the Bitcoin transactions is in fact quite complex. Since it is of a special interest for us, we describe it now in more detail. 
The Bitcoin currency system consists of {\em addresses} and {\em transactions} between them.  
An address is simply a public key $\pk$\footnote{Technically an address is a {\em cryptographic hash} of $\pk$.  
In our informal description we decided to assume that it is simply $\pk$. 
This is done only to keep the exposition as simple as possible, as it improves the readability of the transaction scripts later in the paper.}.
Normally every such key has a corresponding private key $\sk$ known only to one user, which is an {\em owner} of this address.  
The private key is used for signing the transactions, and the public key is used for verifying the signatures.
Each user of the system needs to know at least one private key of some address, but this is simple to achieve, since the pairs $(\sk,\pk)$ can be easily generated 
offline.

\vspace{-.3truecm}

\subsubsection{Simplified version} We first describe a simplified version of the system and then show how to extend it to obtain the description of the real Bitcoin. 
Let $A=(A.\sk,A.\pk)$ be a key pair. In our simplified view a transaction describing the fact that an amount $v$ (called the {\em value} of a transaction) is transferred from an address $A.\pk$ to an address $B.\pk$ has the following form 
$T_{x} = (y,B.\pk,v,\sig_{A}(y, B.\pk,v)),$
where 
$y$ is an index of a previous transaction $T_{y}$.  
We say that $B.\pk$ is the recipient of $T_{x}$, and that the transaction $T_{y}$ is an {\em input} of the transaction $T_{x}$, or that it is {\em redeemed} by this transaction (or redeemed by the address $B.\pk$). 
More precisely, the meaning of $T_{x}$ is that the amount $v$ of money transferred to $A.\pk$ in transaction $T_{y}$ is transferred further to $B.\pk$.   
The transaction is valid only if (1) $A.\pk$ was a recipient of the transaction $T_{y}$, (2) the value of $T_{y}$ was at least $v$ (the difference between the value of $T_{y}$ and $v$ is called the {\em transaction fee}), (3) the transaction $T_{y}$ has not been redeemed earlier, and (4) the signature of $A$ is correct. 
Clearly all of these conditions can be verified publicly.

The first important generalization of this simplified system is that a transaction can have several ``inputs'' meaning that it can accumulate money from several past transactions $T_{y_{1}},\ldots,$ $T_{y_{\ell}}$. 
Let $A_{1},\ldots,A_{\ell}$ be the respective key pairs of the recipients of those transactions.  
Then a multiple-input transaction has the following form: 
$T_{x} = (y_{1},\ldots,y_{\ell},B.\pk, v,\sig_{A_{1}}(y_{1}, B.\pk,$ $v), \ldots, 
\sig_{A_{\ell}}(y_{\ell},B.\pk,v)),$
and the result of it is that $B.\pk$ gets the amount $v$, provided it is at most equal to the sum of the values of transactions $T_{y_{1}},\ldots,T_{y_{\ell}}$.  
This happens only if {\em none} of these transactions has been redeemed before, and {\em all} the signatures are valid.

Moreover, each transaction can have a {\em time lock} $t$ that tells at what time 
in the future the transaction becomes valid. The lock-time $t$ can refer either to a measure called the ``block index'' or to the real physical time. In this paper we only consider the latter type of time-locks. In this case we have 
$T_{x} = (y_{1},\ldots,y_{\ell},B.\pk,v,t,\sig_{A_{1}}(y_{1},B.\pk,v,t),$ 
$\ldots,\sig_{A_{\ell}}(y_{\ell},B.\pk,v,t)).$
Such a transaction becomes valid only if time $t$ is reached and if none of the transactions $T_{y_{1}},\ldots,T_{y_{\ell}}$ has been redeemed by that time (otherwise it is discarded).  
Each transaction can also have several outputs, which is a way to divide money between several users and to divide transactions with large value into smaller
portions. 
We ignore this fact in our description since we will not use it in our protocols.

\vspace{-.3truecm}

\subsubsection{More detailed version} 
The real Bitcoin system is significantly more sophisticated than what is described above.  First of all, there are some syntactic differences, the most important being that each transaction $T_{x}$ is identified not by its index, but by its hash $H(T_{x})$.  Hence, from now on we will assume that $x = H(T_{x})$.

The main difference is, however, that in the real Bitcoin the users have much more flexibility in defining the condition on how the transaction $T_{x}$ can be redeemed. Consider for a moment the simplest transactions where there is just one input and no time-locks.
Recall that in the simplified system described above, in order to redeem a transaction, its recipient $A.\pk$ had to produce another transaction $T_{x}$ signed with his private key $A.\sk$.  In the real Bitcoin this is generalized as follows: each transaction $T_{y}$ comes with a description of a function (\emph{output-script}) $\pi_{y}$ whose output is Boolean.  The transaction $T_{x}$ redeeming the transaction $T_{y}$ is valid if $\pi_{y}$ evaluates to $\true$ on input $T_{x}$. Of course, one example of $\pi_{y}$ is a function that treats $T_{x}$ as a pair (a message $m_{x}$, a signature $\sigma_x$), and checks if $\sigma_{x}$ is a valid signature on $m_{x}$ with respect to the public key $A.\pk$.  However, much more general functions $\pi_{y}$ are possible.  Going further into details, a transaction looks as follows: 
$T_{x} = (y,\pi_{x},v,\sigma_{x}),$
where $\BODY{T_{x}} = (y,\pi_{x},v)$ is called the {\em body}\footnote{In the original Bitcoin documentation this is called ``simplified $T_{x}$''. Following our earlier work \cite{ADMM13,ADMM13b,ADMM13c} we chosen to rename it to ``body'' since we find the original terminology slightly misleading.} of $T_{x}$ and $\sigma_{x}$ is a ``witness'' that is used to make the script $\pi_{y}$ evaluate to $\true$ on $T_{x}$ (in the simplest case $\sigma_{x}$ is a signature on $\BODY{T_{x}}$).
The scripts are written in the Bitcoin scripting language \cite{Script}, which is stack-based and similar to the Forth programming language. It is on purpose not Turing-complete  (there are no loops in it), since the scripts need to evaluate in (short) finite time. It provides basic arithmetical operations on numbers, operations on stack, if-then-else statements and some cryptographic functions like calculating hash function or verifying a signature.

\begin{wrapfigure}{r}{0.6\textwidth}
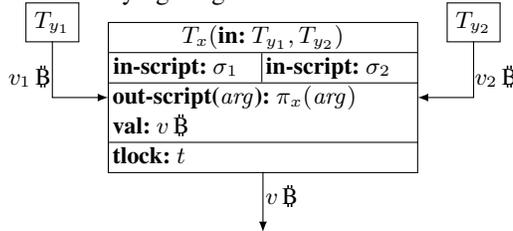

\vspace{-1.3truecm}
\graphC{
  \transTwoInWithScripts[$t$]{0,4}{4}
    {tx}{$T_x(\inp~T_{y_{1}},T_{y_{2}})$}
    {\inScripts
      {$\sigma_{1}$}
      {$\sigma_{2}$}
    }
    {$\argu$}
    {$\pi_{x}(\argu)$}
    {$v\,\BTC$};

  \node[rectangle] (ty1) at (-2.8,5)
  {\fbox{$T_{y_1}$}};
  
  \node[rectangle] (ty2) at (2.8,5)
  {\fbox{$T_{y_2}$}};
  
  \draw [->] ([xshift=-0cm, yshift=-0.0cm]ty1.south) |-
    ([xshift=0.0cm, yshift=0.0cm]tx.west)
    node[midway, xshift=-0.3cm, yshift=0.28cm]{$v_1\,\BTC$};
    
  \draw [->] ([xshift=-0cm, yshift=-0.0cm]ty2.south) |-
    ([xshift=-0.0cm, yshift=0.0cm]tx.east)
    node[midway, xshift=0.3cm, yshift=0.28cm]{$v_2\,\BTC$};
    
  \draw [->] ([xshift=0cm, yshift=0.0cm]tx.south) --
    ++ (0, -0.8)
    node[midway, xshift=0.25cm, yshift=0.08cm]{$v\,\BTC$};
}
\vspace{-.8truecm}\caption{A graphical representation of a transaction $T_{x} = (y_{1},y_{2},\pi_{x},v,t,\sigma_{1},\sigma_{2})$.}\vspace{-1.truecm}\label{fig:gra}
\end{wrapfigure}

The generalization to the mul\-tiple-input transactions with time-locks is straightforward: a transaction has a form:
$T_{x} = (y_{1},\ldots,y_{\ell},$ $\pi_{x},v,t,\sigma_{1},\ldots,\sigma_{\ell}),$
where the body $\BODY{T_{x}}$ is equal to $(y_{1},\ldots,y_{\ell},\pi_{x},v,t),$ and it is valid if (1)  time $t$ is reached, (2) {\em every} $\pi_{i}(\BODY{T_{x}},\sigma_{i})$ evaluates to $\true$, where each $\pi_{i}$ is the output script of the transaction $T_{y_{i}}$, and (3) none of these transactions has been redeemed before.
We will present the transactions as boxes.  
The redeeming of transactions will be indicated with arrows (the arrows will be labelled with the transaction values). An example of a graphical representation of a transaction is depicted in Fig.~\ref{fig:gra}.

The transactions where the input script is a signature, and the output script is a verification algorithm are the most common type of transactions.  We will call them {\em standard transactions}. Currently some miners accept only such transactions,
due to the fact that writing more advanced scripts is hard and error-prone, and anyway the vast majority of users does not use such advanced features of Bitcoin.  Fortunately, there exist other miners that do accept the non-standard (also called {\em strange}) transactions, one example being a big mining pool\footnote{Mining pools are coalitions of miners that perform their work jointly and share the profits.} called {\em Eligius} (that mines a new block on average once per hour). 
We also believe that in the future accepting the general transaction will become standard, maybe at a cost of a slightly increased fee.  Actually, popularizing the Bitcoin contracts, by making them safer to use, is one of the goals of this work.  

\up\up\up 



\section{Modeling the Bitcoin}\label{sec:mod}
  
To reason formally about the security of the contracts we need to describe the attack model that corresponds to the  Bitcoin system.  The model used in \cite{ADMM13,ADMM13b} was inspired by the approach used in the complexity-based cryptography. This way of modeling  protocols, although very powerful, is not well-suited for the automatic verification of cryptographic protocols. 
In this section we present the approach used in this paper, based on timed-automata, using the syntax of the \UPPAAL{} model checker.  

In our model each party executing the protocol is modeled as a timed automaton
with a structure assigned to it that describes the party's knowledge.
States in the automaton describe which part of the protocol the party is performing.
The transitions in the automaton contain conditions, which have to be satisfied for a transition to be taken
and actions, which are performed whenever a transition is taken.

The communication between the parties may be modeled in a number of various ways.
It is possible to use synchronization on channels offered by \UPPAAL~and shared variables representing
data being sent between the parties.
In all protocols verified by us, the only messages exchanged by the parties were signatures.
We decided to model the communication indirectly using shared variables ---
each party keeps the set of known signatures, and whenever a sender wants to send
a signature, he simply adds it to the recipient's set.












The central decision that needs to be made is how to model the knowledge of the honest parties and the adversary.  
Our representation of knowledge is symbolic and based on Dolev-Yao model \cite{DY}, and
hence we assume that the cryptographic primitives are perfectly secure.
In our case it means, for example, that it is not possible to forge a signature without the knowledge of
the corresponding private key, and this knowledge can be modeled in a ``binary'' way: either an adversary knows the key, or not.
The hash functions will be modeled as random oracles, which in  particular implies that they are collision-resilient and hard to invert. 
We also assume that there exists a secure and authenticated channel between the parties,
which can be easily achieved using the public key cryptography.
Moreover, we assume that there is a fixed set of variables denoting the private/public pairs of Bitcoin keys.
A Bitcoin protocol can also involve secret strings known only to some parties (like e.g. a string $s$ in the commitment protocol in Appendix~\ref{app:commit}),
we assume that there is a fixed set of variables denoting such strings.
For each private key and each secret string there is a subset of parties, which know them,
but all public keys and hashes of all secret strings are known to all the parties (if this is not the case then they
can be broadcast by parties knowing them).

A block chain is modelled using a shared variable (denoted \BC) keeping the status of all transactions
and a special automaton, which is responsible for maintaining the state of \BC{} (e.g. confirming transactions).


In the following sections we describe our model in more details.

\subsection{The keys, the secret strings, and the signatures}\label{sec:kss}
\begin{wrapfigure}{R}{0.31\textwidth}
\vspace{-.8truecm}
\begin{framed}
  {\footnotesize{\tt
typedef struct \{
\begin{tabular}{rl}
	\UPPB{Key} & key;\\
	\UPPB{TxId} &tx\_num;\\
	\UPPB{Nonce} &input\_nonce;\\
\end{tabular}\\
\} \UPPB{Signature};
}}
\end{framed}\vspace{-.2truecm}
  \caption{The signatures type.}\label{fig:Sig}\vspace{-.8truecm}
\end{wrapfigure}

We assume that the number of the key pairs in the protocol is known in advance and constant.
Therefore, key pairs will be simply referred by consecutive natural numbers
(type \UPPB{Key} is defined as an integer from a given range).
Secret strings are modelled in the same way.
As already mentioned we assume that all public keys and hashes of all secrets are known to all parties.

Moreover, we need to model the signatures over transactions (one reason is that they are exchanged by the parties in some protocols).
They are modelled by structures containing a transaction being signed, the key used to compute a signature
and an \UPP{input\_nonce}, which is related to the issue of transaction malleability
and described in Appendix~\ref{par:mal}.

\subsection{The transactions}\label{sec:transactions}

We assume that all transactions that can be created by the honest parties during the execution of the protocol
comes from a set, which is known in advance and of size $T$.
Additionally, the adversary can create his own transactions.
As explained later (see Sec.~\ref{sec:adversary} below) we can upper-bound the number
of adversarial transactions by $T$.
Hence the total upper bound on the number of transactions is $2 T$.

\begin{wrapfigure}{R}{0.43\textwidth}
\vspace{-.8truecm}
\begin{framed}
  {\footnotesize{\tt
typedef struct \{
\begin{tabular}{rl}
	\UPPB{TxId} & num;\\
	\UPPB{TxId} &input;\\
	\UPPB{int} &value;\\
	\UPPB{int} &timelock;\\
	\UPPB{bool} &timelock\_passed;\\
	\UPPB{Status} &status;\\
	\UPPB{Nonce} &nonce;\\
	\UPPB{bool} &reveals\_secret;\\
	\UPPB{Secret} & secret\_revealed;\\
	\UPPB{OutputScript} &out\_script;\\
\end{tabular}\\
\} \UPPB{Tx};
}}
\end{framed}\vspace{-.2truecm}

  \caption{The transactions type}\label{fig:Tx}\vspace{-.8truecm}
\end{wrapfigure}

For simplicity we refer the transactions using fixed identifiers instead of their hashes.
We can do this because we know all the transactions, which can be broadcast in advance (compare Sec.\ref{sec:adversary} and Appendix \ref{par:mal} for further discussion).
A single-input and single-output transaction is a variable of a record type \UPPB{Tx} defined in Fig.~\ref{fig:Tx}.
The \UPP{num} field is the identifier of the transaction, and the \UPP{input} field is the identifier of its input transaction.
The \UPP{value} field denotes the value of the transaction (in $\BTC$). The \UPP{timelock} field indicates the time lock of the transaction, and the \UPP{timelock\_passed} is a boolean field indicating whether the \UPP{timelock} has passed.

\noindent
The \UPP{status} field is of a type \UPPB{Status} that contains following values: \UNSENT{} (indicating that transaction has not yet been sent to the block chain),
\SENT{} (the transaction has been sent to the block chain and is waiting to be confirmed), \CONFIRMED{} (the transaction is confirmed on the block chain, but not spent),
\SPENT{}  (the transaction is confirmed and spent), and  \CANCELLED{} (the transaction was sent to the block chain,
but while it was waiting for being included in the block chain its input was redeemed by another transaction).


\noindent
The \UPP{out\_script} denotes the output script. 
In case the transaction is standard it simply contains a public key of the recipient of the transaction.
Otherwise, it refers to a hard-coded function which implements the 
output script of this transaction (see Sec.~\ref{sec:blockchain} for more details).

The inputs scripts are modelled only indirectly (the fields \UPP{reveals\_secret} and \UPP{sec\-ret\_revealed}).
More precisely, we only keep information about which secrets are included in the input script
(see e.g. the  $\AR$ protocol in Fig.~\ref{fig:CS}, transaction $\Reveal$).


The above structure can be easily extended to handle multiple inputs and outputs.


\subsection{The parties}\label{sec:parties} 
\begin{wrapfigure}{r}{0.59\textwidth}
\vspace{-1.8truecm}
\begin{framed}
  {\footnotesize{\tt
typedef struct \{
\begin{tabular}{rl}
  \UPPB{bool} & \UPP{know\_key[KEYS\_NUM];}\\
  \UPPB{bool} & \UPP{know\_secret[SECRETS\_NUM];}\\
  \UPPB{int[0,KNOWN\_}\\
  \UPPB{SIGNATURES\_SIZE]} &\UPP{known\_signatures\_size;}\\
  \UPPB{Signature} & \UPP{known\_signatures}\\
  & \UPP{[KNOWN\_SIGNATURES\_SIZE];}\\
  \end{tabular}\\
\} \UPPB{Party};
}}
\end{framed}
  \vspace{-.3truecm}\caption{The parties type.}\label{fig:party}\vspace{-0.9truecm}
\end{wrapfigure}

The parties are modelled by timed automata describing protocols they follow.
States in the automata describe which part of the protocol the party is performing.
The transitions in the automata contain conditions, which have to be satisfied for a transition to be taken
and actions, which are performed whenever a transition is taken. 
An example of such automaton appears in Fig.~\ref{fig:Recipient}.
and is described in more details in Appendix~\ref{app:syntax}.
The adversary is modelled by a special automaton described in Sec.~\ref{sec:adversary}.

Moreover, we need to model the knowledge of the parties (both the honest users and the adversary) in order to be able to decide whether
they can perform a specific action in a particular situation (e.g. compute the input script for a given transaction).
Therefore for each party, we define a structure describing its knowledge. 
More technically: this knowledge is modelled by a record type \UPPB{Party} defined in Fig.~\ref{fig:party}.  


The boolean tables \UPP{know\_key[KEYS\_NUM]} and \UPP{know\_secret[SECRETS\_NUM]} describe the sets of keys and secrets (respectively) known to the party:
\UPP{know \_key[i] =} \UPP{true} if and only if the party  knows the \UPP{i}-th secret key,
and \UPP{know\_secret[i] = true} if and only if the party  knows the \UPP{i}-th secret string.
The integer \UPP{known\_signatures\_size} describes the number of the additional signatures known to the party (i.e. received from other parties during the protocol),
and the array \UPP{known\_signatures} contains these signatures.

\subsection{The adversary}\label{sec:adversary}
The real-life Bitcoin adversary can create an arbitrary number of transactions with arbitrary output scripts, so it is clear that
we need to somehow limit his possibilities, so that the space of possible states is finite and of a reasonable size.
We show that without loss of generality we can consider only scenarios in which an adversary sends to the block chain transactions only from a finite set
depending only on the protocol. 

The knowledge of an adversary is modeled in the similar way to honest parties,
but we do not specify the protocol that he follows.
Instead, we use a generic automaton, which (almost) does not depend on the protocol being verified
and allows to send to the block chain any transaction at any time assuming some conditions are met, e.g.~that the transaction is valid
and that the adversary is able to create its input script.

We observe that the transactions made by the adversary can influence the execution of the protocol only in two ways:
either (1) the transaction is identical to the transaction from the protocol being verified
or (2) the transaction redeems one of the transactions from the protocol.
The reason for above is that honest parties only look for transactions of the specific form (as in the protocol being executed),
so the only thing an adversary can do to influence this process is to create a transaction, which looks like one of these transactions or redeem one of these.
Notice that we do not have to consider transactions with multiple inputs redeeming more than one of the protocol's transactions, because
there is always an interleaving in which the adversary achieves the same result using a number of transactions with single inputs.
The output scripts in the transactions of type (2) do not matter, so we may assume that an adversary
always sends them to one particular key known only to him.

Therefore, without loss of generality we consider only the transactions,
which appear in the protocol being verified or transactions redeeming one of these transactions.
Hence, the total number of transactions, which are modeled in our system is twice as big as the number of transactions in the original protocol.

\begin{figure}[H]
\centering
\includegraphics[scale=0.77]{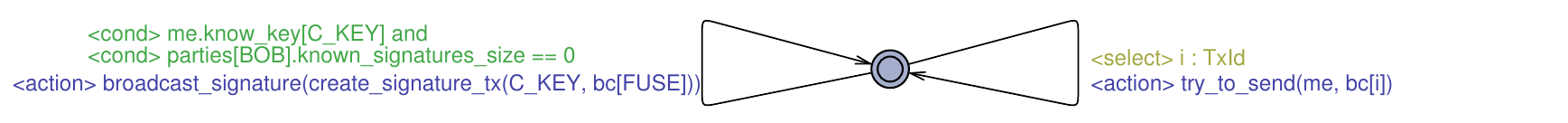}
\caption{The automaton for the Adversary}\label{fig:adv}
\end{figure}

The adversary is then a party, that can send an arbitrary transaction from this set if only he is able to do so (e.g. he is able to
evaluate the input script and the transaction's input is confirmed, but not spent).
If the only actions of the honest parties  is to post transactions on the block chain,
then one can assume that this is also the only thing that the adversary does.
In this case his automaton, denoted \UPP{Adversary} is very simple:
it contains one state and one loop, that simply tries to send an arbitrary transaction from the mentioned set.
This is depicted in Fig.~\ref{fig:adv} on page \pageref{fig:adv} (for a moment ignore the left loop).

In some protocols the parties besides of posting the transactions on the block chain, also exchange messages with each other.
This can be exploited by the adversary, and hence we need to take care of this in our model. This is done by adding more actions in the \UPP{Adversary} automaton.
In our framework, this is done manually. For example in the protocol that we analyze in Sec.~\ref{sec:commit}
Alice sends a signature on the $\Fuse$ transaction (cf.~Step \ref{step:sign}, Fig.~\ref{fig:CS}, Appendix \ref{app:commit}).
This is reflected by the left loop in the \UPP{Adversary} automaton in Fig.~\ref{fig:adv}, which should be read as follows:
if the adversary is able to create a signature on the $Fuse$ transaction and Bob did not receive it yet, then he can send it to Bob.

Of course, our protocols need to be analyzed always ``from the point of view of an honest Alice''
(assuming Bob is controlled by the adversary) and ``from the point of view of an honest Bob'' (assuming Alice is controlled by the adversary).
Therefore, for each party we choose whether to use the automaton describing the protocol executed by the parties or the already mentioned special automaton for an adversary.

%
%

\subsection{The block chain and the notion of time}\label{sec:blockchain}
\begin{figure}[H]
\centering
\includegraphics[scale=0.9]{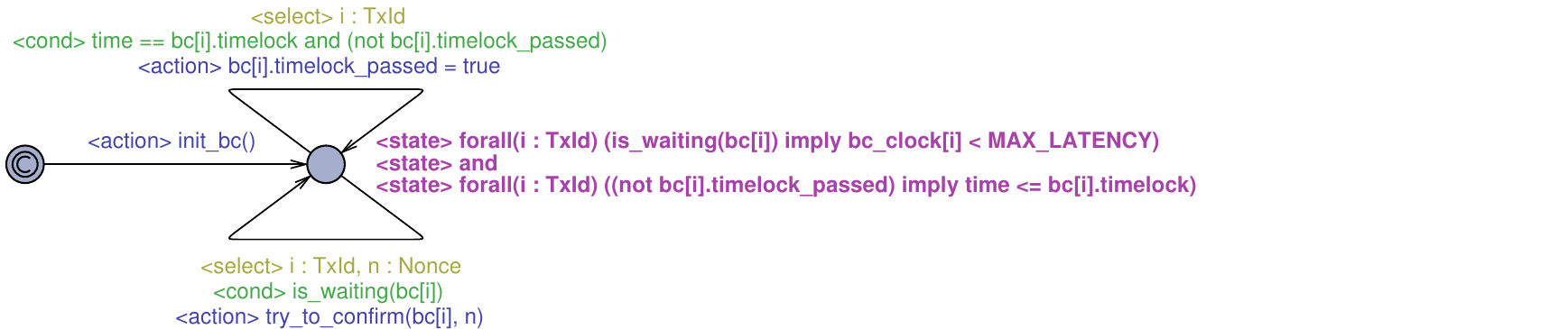}
\caption{The \BCA~automaton}\label{fig:BCA}
\end{figure}

In Bitcoin whenever a party wants to post a transaction on the block chain she broadcasts it over a peer-to-peer network.
In our model this is captured as follows. 
We model the block chain as a shared structure denoted \BC{} containing the information about the status of all the transactions
and a timed automaton denoted \BCA{} (see~Fig.~\ref{fig:BCA}), which is responsible for maintaining the state of \BC.
One of the duties of \BCA{} is ensuring that the transactions which were broadcast are confirmed within appropriate time frames.

%
%

In order to post a transaction \UPP{t} on the block chain,
a party \UPP{p} first runs the \UPP{try\_to\_} \UPP{send(} \UPP{Party\ p,} \UPP{Tx\ t)} function,
which broadcasts the transaction if it is legal.
In particular, the \UPP{can\_send} function checks if (a) the transaction has not been already sent,
(b) all its inputs are confirmed and unredeemed and
(c) a given party \UPP{p} can create the corresponding
input script.
The only non-trivial part is (c) in case of non-standard transactions,
as this check is protocol-dependent.
Therefore,
the exact condition on when the party \UPP{p} can create the appropriate input script, has to be extracted manually from the description of the protocol.
If all these tests succeed, then the function communicates the fact of broadcasting the transaction using the shared structure \BC.

Once a transaction \UPP{t} has been broadcast,
the \BCA{} automaton attempts to include it in the block chain (lower loop in Fig.~\ref{fig:BCA}).
The \BCA{} automaton also enforces that every transaction gets included into the block chain in less than \LATENCY{} time, which is a constant that is defined in the system.
This is done by the invariant on the right state in Fig.~\ref{fig:BCA} that guarantees that every transaction is waiting for confirmation less than \LATENCY{}.

%
%

\subsubsection{Eavesdropping on the network}
The other issue with the block chain is that the peers in the network can see transactions before they are confirmed.
Therefore if a transaction $t$ contains (e.g. in its input script) a secret string $x$ then
an adversary can learn the value of $x$ before $t$ is confirmed and for example use it to create a different
transaction redeeming the input of $t$ (a similar scenario is possible for a two-party lottery protocol from \cite{ADMM13}, which
is only secure in a ``private channel model'').
To take such possibilities into account, broadcasting a transaction results in disclosure
of the secret string $x$, what in our model corresponds to setting appropriate knowledge flags for all parties.

\subsubsection{Malleability of transactions}
\BCA{} automaton is also responsible for choosing the nonces, which imitate the attacks involving the malleability of transactions.
This is described in details in Appendix~\ref{par:mal}.

\out{

\subsection{Security Model}\label{sec:model}

\up

To reason formally about the security we need to describe the attack model that corresponds to the current Bitcoin system. We assume that the parties are connected by an insecure channel and have access to the Bitcoin chain. Let us discuss these two assumptions in detail.  First, recall that our protocol should allow any pair of users on the internet to engage in a protocol.  Hence, we cannot assume that there is any secure connection between the parties (as this would require that they can  verify their identity, which obviously is impossible in general), and therefore any type of a man-in-the middle attack is possible.

The only ``trusted component'' in the system is the Bitcoin chain.  For the sake of simplicity in our model we will ignore the implementation details of it, and simply assume that the parties have access to a trusted third party denoted $\BB$, whose contents is publicly available.  One very important aspect that needs to be addressed are the security properties of the communication channel between the parties and the $\BB$.  Firstly, it is completely reasonable to assume that the parties can verify $\BB$'s authenticity.  In other words: each party can access the current contents of the $\BB$.  In particular, the users can post transactions on the $\BB$.
After a transaction is posted it appears on the $\BB$ (provided it is valid), however it may happen not immediately, and some delay is possible. There is an upper bound $\LATENCY$ on this delay.  This corresponds to an assumption that sooner or later every transaction will appear in some Bitcoin block.  We use this assumption very mildly and e.g.~$\LATENCY = 1$~day is also ok for us (the only price for this is that in such case we have to allow the adversary to delay the termination of the protocol for time $O(\LATENCY)$).  Each transaction posted on the $\BB$ has a time stamp that refers to the moment when it appeared on the $\BB$. 

What is a bit less obvious is how to define privacy of the communication between the parties and the $\BB$, especially the question of the privacy of the writing procedure.  More precisely, the problem is that it is completely unreasonable to assume that a transaction is secret until it appears on the $\BB$ (since the transactions are broadcast between the nodes of the network).  Hence we do not assume it.  This actually poses an additional challenge in designing the protocols because of the problem of the {\em malleability}\footnote{See \url{en.bitcoin.it/wiki/Transaction_Malleability}.} of the transactions. Let us explain it now.
Recall that the transactions are referred to by their hashes.  Suppose a party $\sP$ creates a transaction $T$ and, before posting it on the $\BB$, obtains from some other party $\sP'$ a transaction $T'$ that redeems $T$ (e.g.: $T'$ may be time-locked and serve $\sP$ to redeem $T$ if $\sP'$ misbehaves).  Obviously $T'$ needs to contain (in the signed body) a hash $H(T)$ of $T$.  However, if now $\sP$ posts $T$ then an adversary (allied with malicious $\sP'$) can  produce another transaction $\tT$ whose semantics is the same is $T$, but whose hash is  different (this can be done, e.g., by adding some dummy instructions to the input scripts of $T$).
The adversary can now post $\tT$ on the $\BB$ and, if he is lucky,  $\tT$ will appear on the $\BB$ instead of $T$!  It is possible that in the future versions of the Bitcoin system this issue will be addressed and the transactions will not be malleable.  In Section \ref{sec:2party} we propose a scheme that is secure under this assumption.  We would like to stress that our main schemes (Section \ref{sec:commit} and \ref{sec:lot}) do {\em not} not assume non-malleability of the transactions, and are secure even if the adversary obtains full information on the transactions before they appear on the $\BB$.

We do not need to assume any privacy of the reading procedure, i.e.~each party accesses pattern to $\BB$ can be publicly known.  We assume that the parties have access to a perfect clock and that their internal computation takes no time. The communication between the parties also takes no time, unless the adversary delays it.
These assumptions are made to keep the model as simple as possible, and the security of our protocols does not depend on these assumptions.
In particular we assume that the network is asynchronous and our protocols are also secure if the communication takes some small amount of time.  For simplicity we also assume that the transaction fees are zero.  The extension to non-zero transaction fees is discussed in Section.~\ref{sec:nonzero}.

}


\section{Modeling the Bitcoin-based timed commitment scheme from \cite{ADMM13}}\label{sec:commit}

In this section we describe the ``contract-dependent'' part of our model. 
Our method of modeling and verifying Bitcoin contracts as timed automata is generic and
can be applied to a large class of Bitcoin contracts (and even possibly automatized as described in the paragraph ``Future work'' on page \pageref{sec:future}).
However, it is easier to describe it using a concrete example. 
As a proof-of-concept we constructed the automata corresponding to a very simple contract
called the ``Bitcoin-based timed commitment scheme'' from \cite{ADMM13}.
For the lack of space we only sketch informally what the protocol is supposed to do.
In the protocol one of the parties (called Alice) commits herself to a secret string $s$.
A key difference between this protocol and classic commitment schemes is that Alice is forced to open the commitment (i.e. reveal the string $s$)
until some agreed moment of time (denoted \texttt{PROT\_TIMELOCK}) or pay $1\,\BTC$ to Bob.
The full description can be found in Appendix \ref{app:commit}.
Although the verification of correctness is quite straightforward in this case, we would like to stress
 that our method is applicable to more complicated contracts, like the $\NewSCS$ protocol from \cite{ADMM13} (see Section \ref{sec:news}), for which the correctness
is much less obvious.


\vspace{-.4truecm}
\subsection{The  results of the verification}\label{sec:res}
\vspace{-.25truecm}

Before running the verification procedure in \UPPAAL{} it is necessary to
choose, which parties are honest and which are malicious.

\begin{figure}[H]
\centering
\includegraphics[scale=0.73]{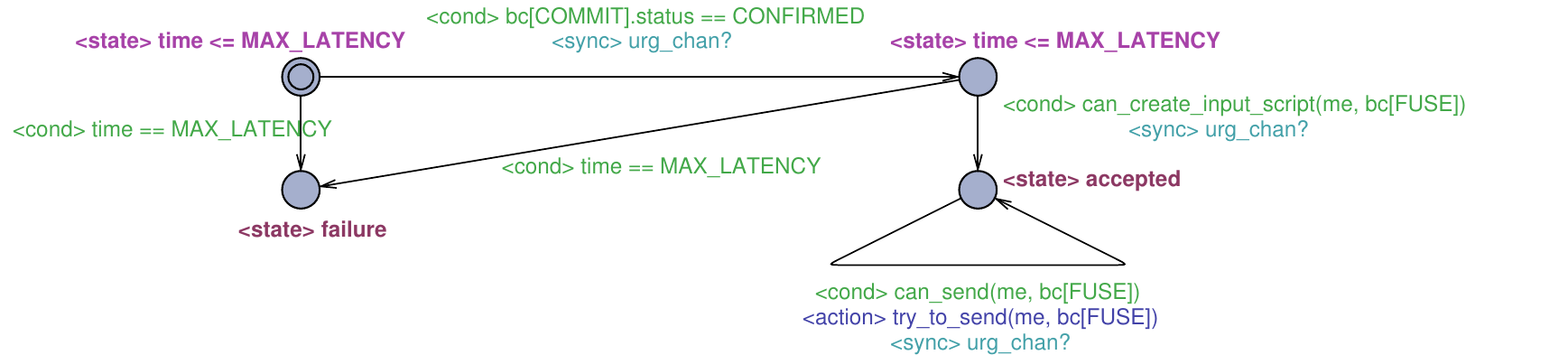}
\caption{The automaton for an honest Bob in timed-commitment scheme}\label{fig:Recipient}
\end{figure}

In \UPPAAL{} it is done by selecting an automaton following the protocol or the
malicious automaton for an adversary described in Sec.~\ref{sec:adversary} for each of the parties.
We started with verification of the security from the point of view of honest Bob.
To this end we used an honest automaton for Bob (see Fig.~\ref{fig:Recipient}) and an adversary automaton described before for Alice (see Fig.~\ref{fig:adv}).

The property that we checked is the following:

\begingroup
\fontsize{8pt}{8pt}\selectfont
\begin{verbatim}
A[] (time >= PROT_TIMELOCK+MAX_LATENCY) imply 
        (hold_bitcoins(parties[BOB]) == 1 or parties[BOB].know_secret[0] 
                                          or BobTA.failure),
\end{verbatim}  
\endgroup
\noindent
which, informally means: ``after time \UPP{PROT\_TIMELOCK+MAX\_LATENCY} one of the following cases takes place: either (a) Bob earned $1\,\BTC$, or (b) Bob knows the committed secret,
or (c) Bob rejected the commitment in the commitment phase''.
This is exactly the security property claimed in \cite{ADMM13}, and hence the verification confirmed our belief that the protocol is secure.
We verified the security from the point of view of Alice in the similar way.

\begin{figure}[H]
\centering
\includegraphics[scale=0.7]{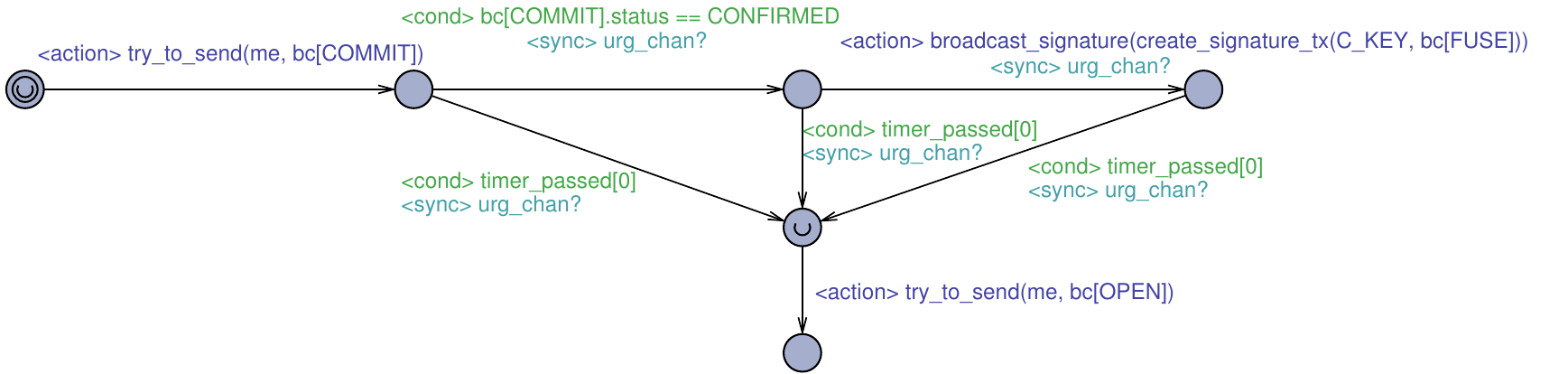}
\caption{The automaton for an honest Alice in timed-commitment scheme}\label{fig:Committer}
\end{figure}

The property we verified means that Alice does not lose any bitcoins in the execution of the protocol (even if Bob is malicious).

As a test we also run the verification procedure on the following two statements:
\begingroup
\fontsize{8pt}{8pt}\selectfont
\begin{verbatim}
       A[] (time >= PROT_TIMELOCK) imply (parties[BOB].know_secret[0])
       A[] (time >= PROT_TIMELOCK) imply (hold_bitcoins(parties[ALICE]) == 1).
\end{verbatim}  
\endgroup
\noindent
The first one states that after time \UPP{PROT\_TIMELOCK} Bob knows the secret 
(which can be not true, if Alice refused to send it).
The second one states that after time \UPP{PROT\_TIME}- \UPP{LOCK} Alice holds 1 \BTC\ (which occurs only if Alice is honest, but not in general).
The \UPPAAL{} model checker confirmed that these properties are violated if one of the parties is malicious, but hold
if both parties follow the protocol (i.e. when honest automata are used for both parties).
Moreover, \UPPAAL{} provides diagnostic traces, which are interleavings of events leading to the violation of the property being tested.
They allow to immediately figure out, why the given property is violated and turned out to be extremely helpful in debugging the automata.

 \vspace{-.3truecm}

\subsection{The $\NewSCS$ protocol from \cite{ADMM13}}\label{sec:news}

We also modeled and verified the Simultaneous Commitment Scheme ($\NewSCS$) protocol from \cite{ADMM13}, which is relatively complicated as it contains $18$ transactions.
To understand it fully the reader should probably look in the~\cite{ADMM13}, but as reference we included the description of these contracts in Appendix \ref{app:scs}.
Informally speaking, the $\NewSCS$ scheme is a protocol that allows two parties, Alice and Bob, to simultaneously commit to their secrets ($s_{A}$ and $s_{B}$, respectively)
in such a way that each commitment is valid only if the other commitment was done correctly. Using \UPPAAL{} we automatically verified the following three conditions,
which are exactly the security statements claimed in \cite{ADMM13b}:
\begin{itemize}
 \item After the execution of the protocol by two honest parties, they both know both secrets and hold the same amount of coins as at the beginning of the protocol, which in \UPPAAL{} syntax was formalized as:
 \begingroup
\fontsize{8pt}{8pt}\selectfont
 \begin{verbatim}  
 A[] (time >=  PROT_TIMELOCK+MAX_LATENCY) imply
    (parties[ALICE].know_secret[SB_SEC] and parties[BOB].know_secret[SA_SEC]
      and hold_bitcoins(parties[ALICE]) == 2
      and hold_bitcoins(parties[BOB]) == 2)
 \end{verbatim}
 \endgroup
 \vspace{-.3truecm}
 (here \UPP{SA\_SEC} and \UPP{SB\_SEC} denote the secrets of Alice and Bob, respectively, and \UPP{2} is the value of the deposit).
\item  An honest Bob cannot lose any coins as a result of the protocol, no matter how the dishonest Alice behaves:
\begingroup
\fontsize{8pt}{8pt}\selectfont
 \begin{verbatim}  
2) A[] (time >= PROT_TIMELOCK) imply hold_bitcoins(parties[BOB]) >= 2
     \end{verbatim}
 \endgroup
  \vspace{-.3truecm}
\item  If an honest Bob did not learn Alice's secret then he gained Alice's deposit as a result of the execution.\begingroup
\fontsize{8pt}{8pt}\selectfont
 \begin{verbatim}  
3) A[] ((time >= PROT_TIMELOCK+2*MAX_LATENCY) imply
         ((parties[ALICE].know_secret[SB_SEC] 
             and !parties[BOB].know_secret[SA_SEC])
         imply hold_bitcoins(parties[BOB]) >= 3))
\end{verbatim}  
\endgroup
\end{itemize}
The analogous guarantees hold for Alice, when Bob is malicious.
The verification of each of the mentioned properties took less than a minute on a dual-core $2.4$ GHz notebook.
We confirmed that the protocol $\NewSCS$ is correct, but there are some implementation details, which are easy to miss and our first
implementation (as an automaton) turned out to contain a bug, which was immediately found due to the verification process and diagnostic traces provided by \UPPAAL.
We describe this in more detail in Appendix \ref{app:bug}.
Moreover \UPPAAL{} turned out to be very helpful in determining the exact time threshold for the time locks, for example we confirmed that the time at which the parties should abort the protocol
claimed in \cite{ADMM13c} ($t-3$\LATENCY) is strict.

These experiments confirmed that the computer aided verification and in particular \UPPAAL{} provides a very good tool for verifying Bitcoin contracts,
especially since it is rather difficult to assess the correctness of Bitcoin contracts by hand,
due to the distributed nature of the block chain and a huge number of possible interleavings.
Therefore, we hope that our paper would encourage designers of complex Bitcoin contracts to make use of computer aided verification 
for checking the correctness of their constructions.

\bibliographystyle{plain}

\bibliography{paper,conf,abbrev_short,crypto}

\def\shortbib{0}
\begin{thebibliography}{10}

\bibitem{icalp1990-AD}
R.~Alur and D.~L. Dill.
\newblock Automata for modeling real-time systems.
\newblock In {\em {ICALP}'90}.

\bibitem{Alur94atheory}
R.~Alur and D.~L. Dill.
\newblock A theory of timed automata.
\newblock {\em Theoretical Computer Science}, 1994.

\bibitem{jacm41(1)-AH}
R.~Alur and T.~A. Henzinger.
\newblock A really temporal logic.
\newblock {\em Journal of the ACM}, 1994.

\bibitem{Times}
T.~Amnell, E.~Fersman, L.~Mokrushin, P.~Pettersson, and W.~Yi.
\newblock {TIMES} - a tool for modelling and implementation of embedded
  systems.
\newblock TACAS '02.

\bibitem{NYT}
Marc Andreessen.
\newblock {Why Bitcoin Matters}, Jan 2013.
\newblock {The New York Times},
  \href{http://dealbook.nytimes.com/2014/01/21/why-bitcoin-matters/}{dealbook.nytimes.com/2014/01/21/why-bitcoin-matters},
  accessed on 26.01.2014.

\bibitem{ADMM13b}
M.~Andrychowicz, S.~Dziembowski, D.~Malinowski, and {\L}.~Mazurek.
\newblock Fair two-party computations via the bitcoin deposits.
\newblock Cryptology ePrint Archive, Report 2013/837, 2013.
\newblock \url{http://eprint.iacr.org/2013/837}, accepted to the 1st Workshop
  on Bitcoin Research.

\bibitem{ADMM13c}
M.~{Andrychowicz}, S.~{Dziembowski}, D.~{Malinowski}, and {\L}.~{Mazurek}.
\newblock {How to deal with malleability of Bitcoin transactions}.
\newblock {\em ArXiv e-prints}, December 2013.

\bibitem{ADMM13}
M.~Andrychowicz, S.~Dziembowski, D.~Malinowski, and {\L}.~Mazurek.
\newblock {Secure Multiparty Computations on Bitcoin}.
\newblock Cryptology ePrint Archive, 2013.
\newblock \url{http://eprint.iacr.org/2013/784}, accepted to the 35th IEEE
  Symposium on Security and Privacy (Oakland) 2014.

\bibitem{UPPAAL}
Gerd Behrmann, Alexandre David, and Kim~G. Larsen.
\newblock A tutorial on uppaal 4.0, 2006.

\bibitem{Petri}
Bernard Berthomieu and Michel Diaz.
\newblock Modeling and verification of time dependent systems using time
  {P}etri nets.
\newblock {\em IEEE Trans. Softw. Eng.}, 17(3):259--273, March 1991.

\bibitem{wiki}
Bitcoin.
\newblock Wiki.
\newblock \href{http://en.bitcoin.it/wiki/}{en.bitcoin.it/wiki/}.

\bibitem{DelfsKnebl}
H.~Delfs and H.~Knebl.
\newblock {\em Introduction to Cryptography: Principles and Applications}.
\newblock Information Security and Cryptography. Springer, 2007.

\bibitem{DY}
D.~Dolev and A.~C. Yao.
\newblock On the security of public key protocols.
\newblock {\em Information Theory, IEEE Transactions on}, 1983.

\bibitem{DolDwoNao00}
Danny Dolev, Cynthia Dwork, and Moni Naor.
\newblock Nonmalleable cryptography.
\newblock {\em {SIAM} Journal on Computing}, 30(2):391--437, 2000.

\bibitem{C:DwoNao92}
C.~Dwork and M.~Naor.
\newblock Pricing via processing or combatting junk mail.
\newblock In {\em CRYPTO}, 1992.

\bibitem{economist1}
The Economist.
\newblock {The Economist explains: How does Bitcoin work?}, Apr 2013.
\newblock
  \href{http://www.economist.com/blogs/economist-explains/2013/04/economist-explains-how-does-bitcoin-work}{www.economist.com/blogs/economist-explains/2013/04/economist-explains-how-does-bitcoin-work},
  accessed on 26.01.2014.

\bibitem{cryptoeprint:2014:076}
G.~Ateniese et~al.
\newblock Certified bitcoins.
\newblock Cryptology ePrint Archive, Report 2014/076.

\bibitem{UPPAAL2}
J.~Bengtsson et~al.
\newblock {UPPAAL} - a tool suite for automatic verification of real-time
  systems.
\newblock In {\em Hybrid Systems}, volume 1066 of {\em Lecture Notes in
  Computer Science}, 1995.

\bibitem{boyen}
S.~Barber et~al.
\newblock Bitter to better - how to make bitcoin a better currency.
\newblock FC'12.

\bibitem{KL}
J.~Katz and Y.~Lindell.
\newblock {\em Introduction to Modern Cryptography (Chapman \& Hall/Crc
  Cryptography and Network Security Series)}.
\newblock Chapman \& Hall/CRC, 2007.

\bibitem{WP}
T.~B. Lee.
\newblock Here's how {B}itcoin charmed {W}ashington.
\newblock
  \href{http://www.washingtonpost.com/blogs/the-switch/wp/2013/11/21/heres-how-bitcoin-charmed-washington/}{www.washingtonpost.com/blogs/the-switch/wp/2013/11/21/heres-how-bitcoin-charmed-washington},
  accessed on 26.01.2014.

\bibitem{CNN}
David~Z. Morris.
\newblock {Bitcoin is not just digital currency. It's Napster for finance}, Jan
  2014.
\newblock CNN Money,
  \href{http://finance.fortune.cnn.com/2014/01/21/bitcoin-platform/}{finance.fortune.cnn.com/2014/01/21/bitcoin-platform},
  accessed on 26.01.2014.

\bibitem{nakamoto}
S.~Nakamoto.
\newblock Bitcoin: A peer-to-peer electronic cash system, 2008.

\bibitem{NicollinS94}
Xavier Nicollin and Joseph Sifakis.
\newblock The algebra of timed processes, atp: Theory and application.
\newblock {\em Inf. Comput.}, 114(1):131--178, 1994.

\bibitem{Forbes}
{R. Cohen}.
\newblock {Global Bitcoin Computing Power Now 256 Times Faster Than Top 500
  Supercomputers, Combined!}
\newblock Forbes,
  \href{http://www.forbes.com/sites/reuvencohen/2013/11/28/global-bitcoin-computing-power-now-256-times-faster-than-top-500-supercomputers-combined/}{www.forbes.com/sites/reuvencohen/2013/11/28/global-bitcoin-computing-power-now-256-times-faster-than-top-500-supercomputers-combined/}.

\bibitem{CSP}
G.~M. Reed and A.~W. Roscoe.
\newblock A timed model for communicating sequential processes.
\newblock {\em Theor. Comput. Sci.}, 58(1-3):249--261, June 1988.

\bibitem{Contr}
Bitcoin wiki.
\newblock Contracts.
\newblock
  \href{http://en.bitcoin.it/wiki/Contracts}{en.bitcoin.it/wiki/Contracts},
  Accessed on 26.01.2014.

\bibitem{Script}
Bitcoin wiki.
\newblock Script.
\newblock \href{https://en.bitcoin.it/wiki/Script}{en.bitcoin.it/wiki/Script},
  Accessed on 26.01.2014.

\bibitem{Mall}
Bitcoin wiki.
\newblock Transaction malleability.
\newblock
  \href{http://en.bitcoin.it/wiki/Transaction_Malleability}{en.bitcoin.it/wiki/Transaction\_Malleability},
  Accessed on 26.01.2014.

\bibitem{PA}
Wang Yi.
\newblock {CCS} + time = an interleaving model for real time systems.
\newblock In {\em Automata, Languages and Programming}, volume 510 of {\em
  Lecture Notes in Computer Science}. 1991.

\bibitem{Kronos}
Sergio Yovine.
\newblock Kronos: a verification tool for real-time systems.
\newblock {\em Journal on Software Tools for Technology Transfer}, 1, October
  1997.

\end{thebibliography}

\appendix
\section{Malleability of transactions}\label{par:mal}
Let us now describe the reason why we introduced the variables of a type \UPPB{Nonce}  in the transaction and the signatures.
The reason is that they are used to model the fact that the transactions  in Bitcoin can be slightly modified without changing its functionality.
Malleability is a very general concept introduced  in cryptography in a seminal paper of Dolev et al.~\cite{DolDwoNao00}. 
It is a (usually) undesired property of the cryptographic schemes that very informally  means that an adversary, after seeing an output of a scheme, can produce another output that is in some non-trivial way ``related'' to the original output.\footnote{For example, the one time pad encryption scheme (see, e.g., \cite{KL}) defined as: $\Enc(K,M) := K \oplus M$ and $\Dec(K,C) := K \oplus C$ (where ``$\oplus$'' denotes the coordinate-wise xor) {\em is} malleable since by negating every bit in $C$ one  obtains a ciphertext $C'$ of a message $M'$ that is equal to a ``negated'' $M$.  Note that this works even if one does not know $K$. Since the one-time pad is perfectly secure, thus the non-malleability is not implied by the standard security of an encryption scheme.} 
As it turns out, malleability of transaction is a feature of Bitcoin, which poses a serious risk for
almost all contracts using time locks.  We now briefly describe this problem. More detailed descriptions can be found in \cite{ADMM13c,Mall}.

Bitcoin transactions are malleable in the following way: given a valid transaction $t$, an adversary is  able 
to create a functionally equivalent and valid transaction $t'$ which has a 
different hash, even if he does not know the secret key used to produce $t$.
Both transactions have the same input transactions, the same output script and differ only in the hash.
Therefore, the following scenario is possible. A transaction $t$ is sent to the block chain, which technically means that it is broadcast over the network.
An adversary can therefore see $t$ and create and broadcast $t'$. If he is lucky then eventually $t'$ becomes included in the block chain, instead of 
$t$. It means that we can not assume the knowledge of the hash of the transaction before it is confirmed.
It poses a serious problem for most of the protocols using time locks, 
because such protocols often need to sign a transaction
$s$ which redeems $t$, {\em before} $t$ is broadcast. 
Such transaction $s$ contains a hash of $t$ and in case when
$t'$ is included in the block chain instead of $t$, the transaction $s$ becomes invalid.

In our model the transactions are addressed by fixed identification numbers instead of hashes, so
a special technique should be applied to take the malleability of transactions into account.
To this end we extend the transaction structure with a ``malleability nonce'', which contains
an integer from a fixed and small set \UPPB{Nonce} (in most cases a set \UPPB{int}\UPP{[0,1]} of size $2$ is enough).
Each transaction, which has not been sent to the block chain has the \UPP{nonce}
field equal to $0$, which indicates that
the transaction was not modified.
Whenever a transaction is being confirmed, its malleability nonce is set to a random value.
Moreover, the structure describing the signature on the transaction $t$ which redeems
$s$ (see Sec.~\ref{sec:kss}) contains
the assumed malleability nonce of $s$ (set when the transaction $t$ was being signed)
and is considered valid only if this nonce matches the real nonce of the transaction $s$.
Therefore, if the transaction $s$ was confirmed after
the signature structure was created, the signature may or may not become invalidated, what catches
the issue of malleability in real Bitcoin network.

The above solution with ``malleability nonces'' makes the assumption that whenever $s$ is modified, then only
the transaction $t$ becomes invalid. In reality it would also influence another transaction $u$
which redeems $t$ and so on. However, in all Bitcoin protocols we are aware of, this is not an issue
as the signatures are computed at most ``one step forward''. It is also not difficult
to extend the model to handle ``malleability nonce of the second (or arbitrary) level'' if necessary.


\newpage

\section{Bitcoin-based timed commitment scheme}\label{app:commit}

\begin{figure}[H]
  \graphC{
    \input{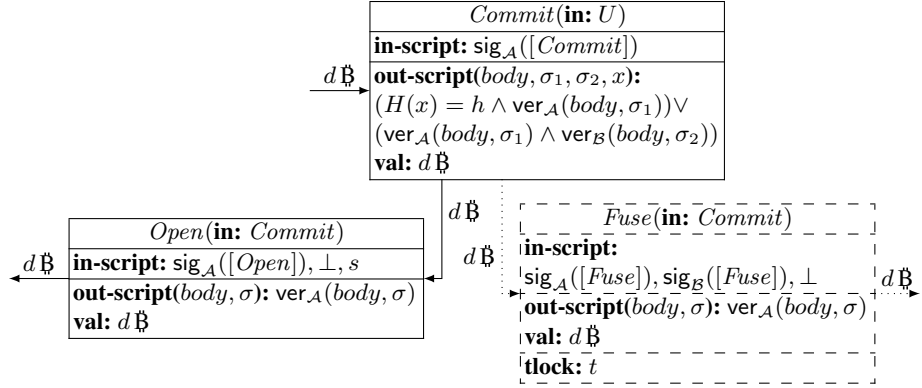}
  }
  
  \vspace{.3truecm}
  
  \desc[0.9]{
    \input{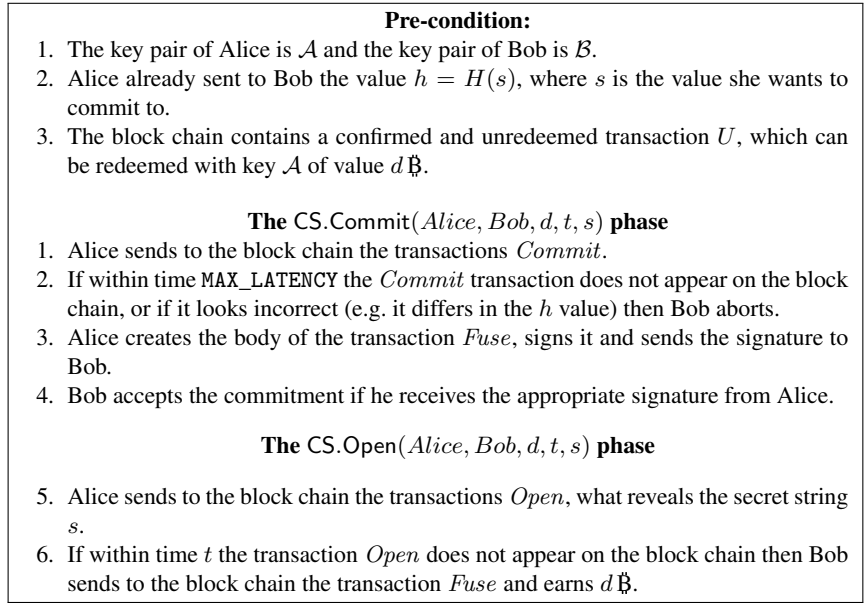}
  }
\caption{The $\AR(Alice, Bob, d, t, s)$ protocol. The scripts' arguments, which are omitted are denoted by $\bot$.}
\label{fig:CS}
\end{figure}

A {\em commitment scheme} is one of the basic cryptographic primitives (see, e.g., \cite{DelfsKnebl}).
It is executed between two parties: Alice and Bob.
In the first phase (called, the {\em commitment phase}) Alice {\em commits} herself to some secret value $s$.
That means that she sends to Bob a value $c = f(s)$ for some agreed  randomized function $f$.
In the second phase (called the {\em opening phase}) Alice sends to Bob $s$ plus some extra information.
%

From the commitment scheme we require two properties --- {\em hiding} and {\em binding}.
These properties will be satisfied if we use $f$ as $f(s) = H(s || r)$ where $H$ is a cryptographic hash function and $r$ is a random, fix length padding (if we model $H$ as a random oracle).
In the Bitcoin-based timed commitment scheme this construction is used as a building block (with often used in Bitcoin function SHA-256 as a hash function).
In its description we assume, that the padding was already added and simply use $H$ instead of $f$.

One of the problems with the classical commitment scheme is that there is no way to force Alice to open the commitment and hence reveal the secret.
To remedy this, in \cite{ADMM13} we proposed to use the Bitcoin system to punish financially Alice in case she does not  open the commitment.
This is done as follows.
Assume that Alice wants to commit herself to some secret $s$.
To do it,  she creates and sends to the block chain a Bitcoin transaction ($\Commit$) that contains a commitment $f(s)$ and has some value $d\, \BTC$ (this money is called the ``deposit''). This transaction  can be spent in one of two ways: either by revealing the secret (this is an expected action of Alice, she will have to reveal the secret to get her money back), or the signatures of both Alice and Bob (by this Bob can punish Alice).
After the transaction is included in the block chain, Alice creates the $\FuseAR$ transaction --- it spends the $\Commit$ transaction and sends the money to Bob.
Alice sets a time lock $t$ (which forces her to open the commitment by time $t$) adds her signature and sends it to Bob.
Bob accepts the commitment if the transactions are correct (e.g. they have proper values, time lock, and $\FuseAR$ has a good signature) and adds his signature to the $\FuseAR$ transaction.
Otherwise he quits.
Alice should open the commitment (i.e. send to the block chain the transaction $\Reveal$ spending the $\Commit$ transaction; it will contain the secret) before the time lock ends to get back her money.
If she does not do that, then Bob sends the $\FuseAR$ transaction to get agreed amount of bitcoins.  The graph of transactions and a detailed description of the protocol is presented in Fig.~\ref{fig:CS}. We require that this scheme, besides of being binding and hiding, has the following extra properties:

\begin{enumerate}
  \item honest Alice will never lose her money and she will always open the commitment,
  \item if both Alice and Bob are honest, then Bob will accept the commitment,
  \item if honest Bob accepts the commitment, then he will learn the secret or gain agreed value of bitcoins.
\end{enumerate}
We verified these conditions using \UPPAAL~and the model described in this paper (see Sec.~\ref{sec:res}).

\newpage


\section{Simultaneous Commitment scheme from \cite{ADMM13c}}\label{app:scs}

\begin{figure}[H]
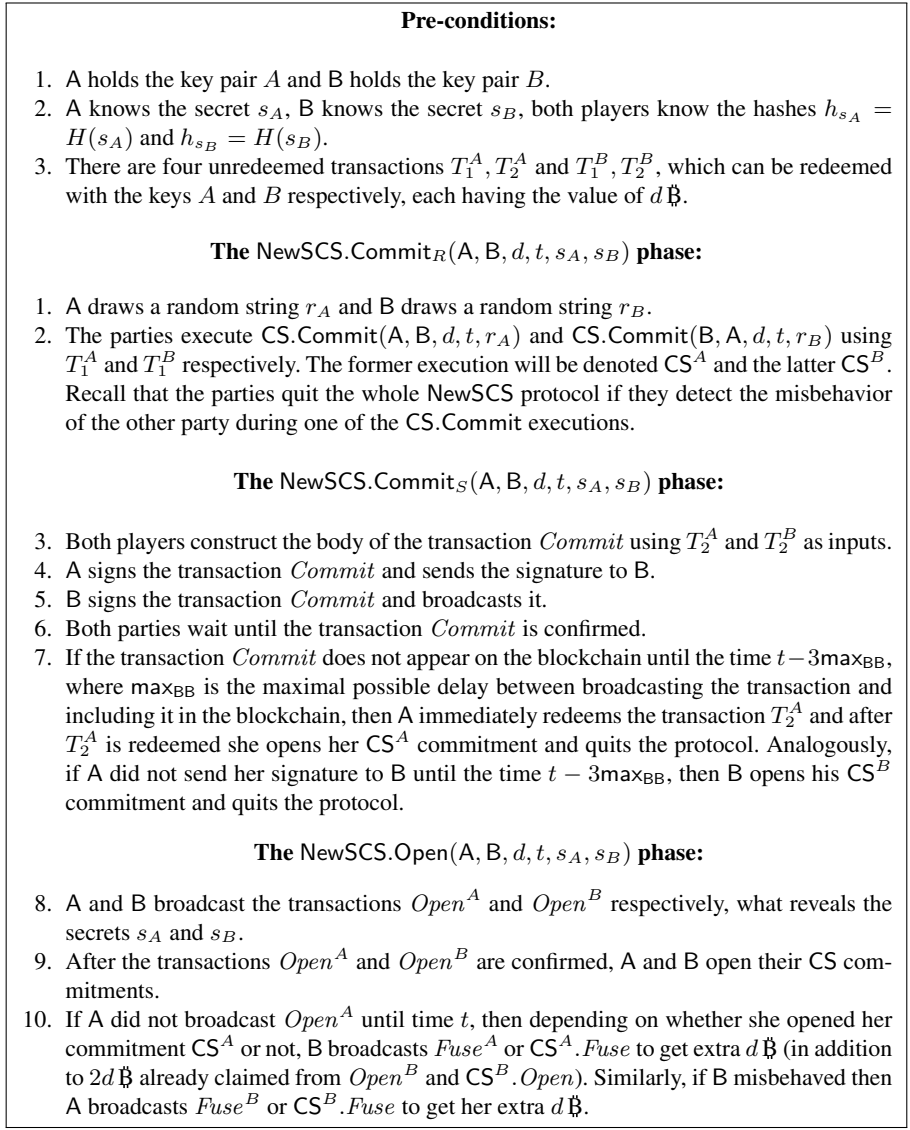

  \desc{
\begin{enumerate}
\item[]
    \begin{center} {\bf Pre-conditions:} \end{center}
\vspace{0.4cm}
  \item $\sA$ holds the key pair $A$ and $\sB$ holds the key pair $B$.
  \item $\sA$ knows the secret $s_A$, $\sB$ knows the secret $s_B$, both players know the hashes $h_{s_A} = H(s_A)$ and $h_{s_B} = H(s_B)$.
  \item There are four unredeemed transactions $T^A_1, T^A_2$ and $T^B_1, T^B_2$, which can be redeemed with the keys $A$ and $B$ respectively, each having the value of $d \, \BTC$. 
\end{enumerate}
    \begin{center} {\bf The $\NSCSCom_R(\sA, \sB,d,t,s_A, s_B)$ phase:} \end{center}
\begin{enumerate}
  \item $\sA$ draws a random string $r_A$ and $\sB$ draws a random string $r_B$.
  \item The parties execute $\CSCom(\sA, \sB, d, t, r_A)$ and $\CSCom(\sB, \sA, d, t, r_B)$ using $T^A_1$ and $T^B_1$ respectively.
 The former execution will be denoted $\ComS^A$ and the latter $\ComS^B$.
 Recall that the parties quit the whole $\NSCS$ protocol if they detect the misbehavior of the other party during one of the
$\CSCom$ executions.
\vspace{0.4cm}
\item[]
    \begin{center} {\bf The $\NSCSCom_S(\sA, \sB,d,t,s_A, s_B)$ phase:} \end{center}
\vspace{0.4cm}
  \item Both players construct the body of the transaction $\CommitSCS$ using $T^A_2$ and $T^B_2$ as inputs.
  \item $\sA$ signs the transaction $\CommitSCS$ and sends the signature to $\sB$.
  \item $\sB$ signs the transaction $\CommitSCS$ and broadcasts it. \label{step:commit}
  \item Both parties wait until the transaction $\CommitSCS$ is confirmed.
  \item If the transaction $\Commit$ does not appear on the blockchain until the time $t-3 \maxBB$, where $\maxBB$ is the maximal possible
  delay between broadcasting the transaction and including it in the blockchain, 
  then $\sA$ immediately redeems the transaction $T^A_2$ and after $T^A_2$ is redeemed she opens her $\ComS^A$ commitment and quits the protocol.
  Analogously, if $\sA$ did not send her signature to $\sB$ until the time $t-3 \maxBB$,
  then $\sB$ opens his $\ComS^B$ commitment and quits the protocol.
\vspace{-0.0cm}
    \begin{center} {\bf The $\NSCSOp(\sA, \sB,d,t,s_A, s_B)$ phase:} \end{center}
\vspace{-0.0cm}
  \item $\sA$ and $\sB$ broadcast the transactions $\Open^A$ and $\Open^B$ respectively, what reveals the secrets $s_A$ and $s_B$.
  \item After the transactions $\Open^A$ and $\Open^B$ are confirmed, $\sA$ and $\sB$ open their $\ComS$ commitments.
  \item If $\sA$ did not broadcast $\Open^A$ until time $t$, then depending on
  whether she opened her commitment $\ComS^A$ or not, $\sB$ broadcasts $\Fuse^A$ or $\ComS^A.\Fuse$ to get extra $d\,\BTC$
  (in addition to $2 d\,\BTC$ already claimed from $\Open^B$ and $\ComS^B.\Open$). 
  Similarly, if $\sB$ misbehaved then $\sA$ broadcasts $\Fuse^B$ or $\ComS^B.\Fuse$
  to get her extra $d\,\BTC$.
\end{enumerate}
  }
\caption{The description of the $\NSCS$ protocol.}
\label{fig:SCS-desc}
\end{figure}

\begin{figure}[H]
 \graphC{
    \input{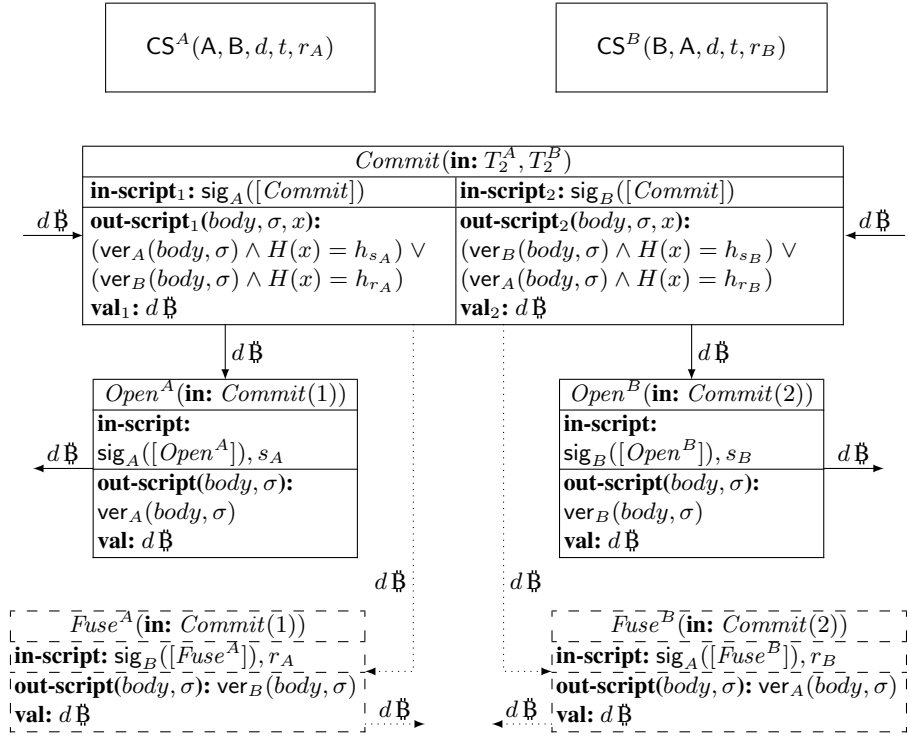}
 }
\caption{The graph of transactions for the $\NSCS$ protocol.
Two boxes labeled with $\AR(\ldots)$ denote the transactions broadcast in the appropriate execution of the Bitcoin-based timed commitment scheme.
$h_x$ denotes the value $H(x)$, but it is used in the output scripts to stress
that the value of the hash is directly included in the transaction (instead of value of $x$ and an application of the hash function).}
\label{fig:SCS-graph}
\end{figure}


\subsection{A bug in the first version of $\NewSCS$}\label{app:bug}
As mentioned in Section~\ref{sec:news}  the protocol $\NewSCS$ is correct, but there are some implementation details, which are easy to miss and our first
implementation turned out to contain a bug, which was immediately found due to verification process.  More precisely, in Step $10$ of the $\NewSCS$ protocol it is said that Bob sends to the block chain one of the two transactions: $\Fuse^A$ or \CS$^A.\Fuse$ and as a result gains $1$ bitcoin, what means that at least one
of these two transactions becomes confirmed. Therefore, our first implementation just tried to send these two transactions.
However, it turns out that there is a strategy of malicious Alice and an interleaving of events in which none of these transactions becomes confirmed and it is necessary
to try to sends the transaction $\Fuse^A$ again after waiting for time \LATENCY.
\UPPAAL~provides diagnostic traces, which allows to easily find bugs like the one mentioned.

\section{\UPPAAL}\label{sec:uppaal-desc}

\subsection{A very short introduction to timed automata and \UPPAAL}

For the lack of space we only sketch the description of \UPPAAL{}~and the semantics of the timed automata (focusing on the most relevant  features).
The reader may  consult \cite{UPPAAL,UPPAAL2} for more information on this topic.  

In \UPPAAL{} the automata are generally finite-state automata equipped with clocks.
\UPPAAL{} system consists of some number of timed automata, clocks and variables of discrete, bounded sets (also user defined, like records) and functions working on variables and clocks.
The state of the system consists of locations of the automata, values of the variables and the clocks evaluation.
The clocks evaluate to positive real numbers, but because of bounded set of possible constraints number of different state
it which the system can be is finite.

The only possible transitions in \UPPAAL~are moving all the clocks forward with the same value or to use some of edges to change the state of some of the automata.
Both kind of transitions are correct only if all \emph{invariants} and \emph{guards} are satisfied.
The invariants are properties of the locations and have to be satisfied each time the system is in this location.
The locations may have also names (used in a verification), be urgent (time cannot pass when any automaton is in such location) or committed (the system immediately has to use an edge outgoing from such location).
Moreover, edges can use selectors to set a local variable to any value of some type.
When an edge is used, it may also run a connected update --- it changes values of variables and resets some clocks.
The pairs of edges may also be synchronized --- in such case they can be used, but only together.
The synchronization may also be urgent --- then it has to be used if only it can be used.
More details and a syntax used in \UPPAAL{} can be found in Appendix \ref{app:syntax}.

\UPPAAL~comes also with a simulator (to e.g. run transitions in a random order) and a verifier.
The verifier checks whether the given properties (written in the simplified version of TCTL --- timed computation tree logic) are satisfied by the  system.


\subsection{\UPPAAL\ syntax}
\label{app:syntax}

\begin{figure}[H]
\centering
\includegraphics[scale=0.73]{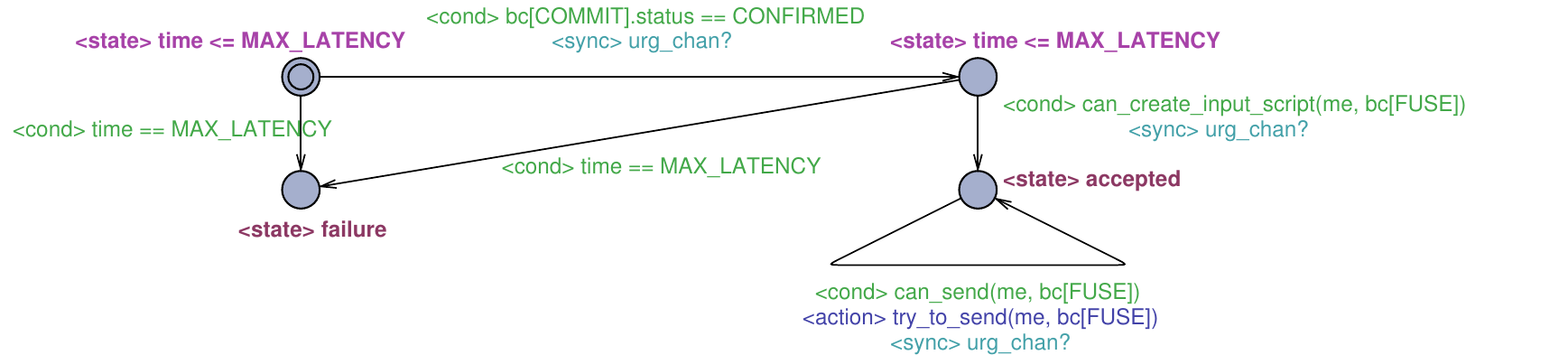}\caption{The automaton for Bob in timed-commitment scheme}\label{fig:bob}
\end{figure}

Let us analyze the \UPPAAL\ syntax on the example of automaton for Bob from a 
timed commitment scheme (Fig.~\ref{fig:bob}).
It has $4$ states: the upper-left is a starting state, the upper-right corresponds to a situation when
the $\Commit$ transaction is confirmed, but the signature on $\Fuse$ from Step.~\ref{step:sign} has not yet been received by Bob.
The lower-left state (denoted \emph{failure}) means that the commitment was not accepted by Bob and the lower-right (denoted \emph{accepted})
state means that Bob accepted the commitment (Step~\ref{step:acc}).

There are $5$ types of labels on the picture\footnote{We color them using the default \UPPAAL{} coloring. We also decided to add special text labels (in the ``<>'' brackets) in the Figures of automata to make the text readable  when printed black and white}:
\begin{description}
  \item[\textcolor{ForestGreen}{\UPPB{<select>}}] 
    These labels are placed on the edges to set a local variable used in following condition and/or action to any value from a specified set,
e.g. \UPP{i : TxId}, \UPP{cond(i)}, \UPP{action(i)} placed on an  edge means that this edge can be used for any \UPP{i} from \UPP{TxId}, provided \UPP{cond(i)} is true and then \UPP{action(i)} is performed.
The mentioned automation for Bob does not have any edges of this kind.
  \item[\textcolor{ForestGreen}{\UPPB{<cond>}}]
    These labels are placed on the edges and they represent conditions (called \emph{guards} in \UPPAAL), which have to be satisfied, e.g. \UPP{bc[COMMIT].status == CON}- \UPP{FIRMED}
    means that this transition can be taken only if the transaction $\Commit$ has been already confirmed and \UPP{can\_create\_input\_script(me, bc[FUSE])}
    checks if Bob is able to create input script for $\Fuse$ transaction, what is equivalent to the fact
    that he has already received the signature from Step.~\ref{step:sign} (Fig.~\ref{fig:CS}).
  \item[\textcolor{Blue}{\UPPB{<action>}}] 
    These labels are placed on the edges and they represent actions (called \emph{updates} in \UPPAAL), i.e.~functions which are called whenever a transition is taken.
    In our example there is just one action \UPP{try\_to\_send(me, bc[FUSE])}, which checks whether it is possible to send the given transaction in the current state
    (e.g.~it was not sent earlier, its inputs are confirmed, but not spent, the current party is able to evaluate the corresponding input scripts) and
    changes the state of the transaction to \SENT~if all conditions are met.
  \item[\textcolor{Cerulean}{\UPPB{<state>}}] 
    These labels, placed on the states, represent names of the states (\emph{failure} and \emph{accepted})
    and \emph{invariants} specifying conditions, which has to be satisfied in the given state.
    Execution in which the invariant is not satisfied are ignored by \UPPAAL.
    Therefore a node with an invariant \UPP{time <= MAX\_LATENCY} and an outgoing edge with a guard \UPP{time ==} \UPP{MAX\_LATENCY}
    guarantee that the edge will be taken exactly at time \LATENCY{} (Moreover it makes impossible to enter the state after time \LATENCY,
    but it is not important in our example.)
  \item[\textcolor{Fuchsia}{\UPPB{<sync>}}] 
    These labels, placed on the edges, represent synchronization channels.
    Here they are used for a special purpose of obtaining urgent transitions.
    This is the technical trick described in Sec.~\ref{sec:helper}, but for understanding this picture it is enough to know
    that the automaton is not allowed to wait whenever there is an edge available with the label \UPP{urg\_chan?}.
\end{description}
%


\subsection{Helper automaton}\label{sec:helper}

A transition is called \emph{urgent}, when it has to be taken immediately whenever it is possible.
More precisely, the time cannot pass, whenever there is an automaton with an urgent transaction available (i.e. with a satisfied guard).
\UPPAAL~does not provide urgent transitions, but there exists a simple workaround, described e.g. in \cite{UPPAAL}.
The solution is based on the so-called urgent {\em channels}. Here, the ``urgency'' means that the time cannot pass whenever there is an automaton
with an available transition synchronizing on an urgent channel (availability means that in particular there is another automaton, which can synchronize
on the same channel). Therefore, it is enough to mark edges we would like to be urgent with a synchronization label \UPP{urg\_chan?}
and create a special automaton (called \Helper~in our implementation) with one state and a loop with label \UPP{urg\_chan!} for
some urgent channel \UPP{urg\_chan}.

\begin{figure}[H]
\centering
\includegraphics[scale=0.97]{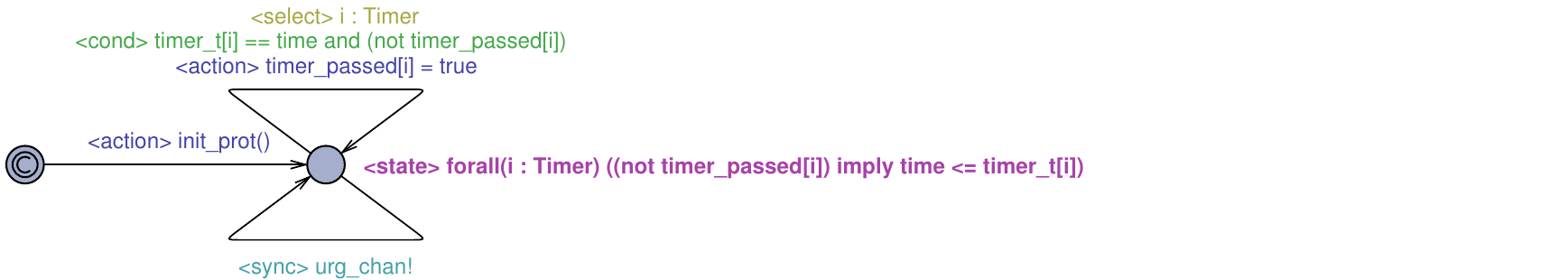}
\vspace{-.2truecm}
\caption{The \UPP{Helper}~automaton}\label{fig:helper}
\end{figure}

\UPPAAL~does not allow to put guards involving clocks on edges with synchronization on urgent channels, so another workaround
is needed to achieve this functionality.
The simple solution is to check on the edge a value of some boolean shared variable and make sure that the value
of this variable is always the same as the value of the clock condition we would like to have on the edge.
The correct value of the shared variable can be maintained by another loop in the \Helper~automaton.
The \Helper~automaton is presented in Fig.~\ref{fig:helper}.

\end{document}